\documentclass[showpacs,preprintnumbers,amsmath,amssymb,pre]{revtex4}

\usepackage{graphicx}

\newcommand{\dd}{{\rm d}}
\newcommand{\mch}{\mathcal{H}}

\begin{document}

\title{Roughness of moving elastic lines~--~crack and wetting fronts}
\author{E. Katzav, M. Adda-Bedia, M. Ben Amar and A. Boudaoud}
\affiliation{Laboratoire de Physique Statistique de l'Ecole Normale Sup\'erieure, CNRS UMR 8550,\\
24 rue Lhomond, 75231 Paris Cedex 05, France.}
\date{\today}

\begin{abstract}
We investigate propagating fronts in disordered media that belong to the universality class of wetting contact lines and planar tensile crack fronts. We derive from first principles their nonlinear equations of motion, using the generalized Griffith criterion for crack fronts and three standard mobility laws for contact lines. Then we study their roughness using the self-consistent expansion. When neglecting the irreversibility of fracture and wetting processes, we find a possible dynamic rough phase with a roughness exponent of $\zeta=1/2$ and a dynamic exponent of $z=2$. When including the irreversibility, we conclude that the front propagation can become history dependent, and thus we consider the value $\zeta=1/2$ as a lower bound for the roughness exponent. Interestingly, for propagating contact line in wetting, where irreversibility is weaker than in fracture, the experimental results are close to $0.5$, while for fracture the reported values of $0.55$--$0.65$ are higher.
\end{abstract}

\pacs{62.20.Mk,68.08.Bc,05.40.-a}

\maketitle

\section{Introduction}

The propagation of fronts or lines in disordered media occurs in a number of fields~\cite{Fisher98}. In particular, the propagation of cracks in heterogeneous materials and the wetting of disordered substrates involve lines with long-range elasticity. These two cases have attracted much attention over the past few years due to the difficulty to match theoretical predictions with experimental measurements. A commonly studied quantity is the so-called roughness exponent $\zeta$, which measures the roughness of the elastic line. For crack propagation, three different exponents must be distinguished~\cite{bouchaud,RafiFrac}: one describing the roughness in the direction perpendicular to the crack propagation, a second the roughness in the direction of the propagation, and a third one (which will interest us in the following) describing the in-plane roughness of the crack front during its propagation through the material. For the spreading/retraction of a liquid wedge on a flat substrate, the contact line has only one roughness exponent. If $h(x,t)$ stands for the transverse perturbation of the position of the line, the linearized equation of motion takes the non-dimensional form
\begin{equation}
\frac{\partial h}{\partial t}(x,t)= \mathcal{PV}\int\frac{h(x',t)-h(x,t)}{(x'-x)^2}\frac{\mathrm{d}x'}{2\pi} + \mathrm{noise} \,,
\label{eq:linear}
\end{equation}
for both tensile cracks~\cite{GaoRice89,Ramanathan97,Ramanathan98} and contact-lines~\cite{JdG,Pomeau,Robbins}. Here and elsewhere $\mathcal{PV}$ stands for the principal value of the integral. Solutions of stochastic growth models exhibiting scaling behavior are described by the time dependent correlation function
\begin{equation}
\left\langle \left[ h(x,t) - h(x',t') \right]^2 \right\rangle = |x-x'|^{2\zeta} f\left( \frac{|x-x'|}{|t-t'|^{z}}\right)\,,
\label{scaling}
\end{equation}
where $\zeta$ is the roughness exponent of the interface and $z$ is the dynamic exponent. From now on, the brackets $\left\langle\cdots\right\rangle$ denote average over disorder. However, for a medium with uncorrelated heterogeneities, the roughness exponent corresponding to Eq.~(\ref{eq:linear}) was found~\cite{Ramanathan97,Ramanathan98,Rosso} to be $\zeta=0.39$, whereas experiments yield $\zeta=0.55$ to $0.65$ for tensile cracks~\cite{Daguier,Schmittbuhl,Santucci06} and slightly more than $\zeta=0.5$ for contact lines~\cite{Prevost,Moulinet}.

To resolve this contradiction, besides accounting for additional physical effects (e.g.~\cite{SHB}), it was proposed~\cite{Golestanian,LeDoussal,PRE,EPL,Procaccia} to include quadratic terms in $h(x,t)$ in the equation of motion. However, different roughness exponents were predicted using the one-loop renormalisation group--$\zeta=0.45$~\cite{LeDoussal}, the self-consistent expansion--$\zeta=1/2$~\cite{EPL}, or numerical simulations--$\zeta=0.37$~\cite{Procaccia}. When looking at the exact form of the equations used in these studies, it turns out that nonlinear terms have different forms. This is the first motivation of the present study, i.e. establishing the nonlinear equations of motion on firm grounds. In addition, experiments~\cite{Daguier,Schmittbuhl,Santucci06,Prevost,Moulinet} are mostly interpreted~\cite{Ramanathan97,Ramanathan98,Rosso,LeDoussal} in terms of the depinning transition of the front. Underlying assumptions are that the driving force is approximately constant, and that the statistical properties of the shape of the front are imposed at its depinning threshold. In fact, the front can achieve high velocities in experiments, well above the depinning threshold. Besides, the recent numerical simulations \cite{Procaccia} suggest that the roughness at depinning is not affected by nonlinear terms in the equation of motion. Therefore, we will investigate the roughness of a front advancing at finite velocity.

Here we study the dynamic roughness of lines advancing according to the nonlinear equations that we obtained. We find a rough phase characterized by an exponent $\zeta=1/2$, close to that observed for contact lines in wetting and slightly below those measured for crack fronts. We suggest that the experimental roughness results from the freezing of an advancing rough front, so that it should be not only close to $\zeta=1/2$ predicted here, but it should also be history dependent, a fact that could be checked in experiments. We will use the Self-Consistent Expansion (SCE), a method developed by Schwartz and Edwards \cite{SCE,TCN04} and applied successfully to the Kardar Parisi-Zhang (KPZ) equation \cite{KPZ}. The method gained much credit by being able to give reasonable predictions for the KPZ scaling exponents in the strong-coupling phase above one dimension where many renormalization group (RG) approaches failed \cite{Wiese98}. Another point which is especially relevant for our purpose is that for a family of models with long-range interactions of the kind treated here, SCE reproduced exact one-dimensional results while RG failed to do so \cite{Nonlocal}.

The article is organized as follows. In section~II, we derive from first principles the nonlinear equations of motion for crack fronts, and for the three available models for contact lines. We set all equations into a single framework and then introduce the annealed noise approximation corresponding to a line advancing at finite velocity. In section~III, we describe the Self-Consistent Expansion and apply it to the nonlinear equation of motion, in order to investigate the possible phases and their roughness. Finally, in section~IV we compare our results with previous works and discuss the relevance of our approach to experiments.

\section{Equation of motion of an elastic line}

\subsection{Fracture}

\subsubsection{The stress-intensity factor}

\begin{figure}[ht]
\centerline{\includegraphics[width=7cm]{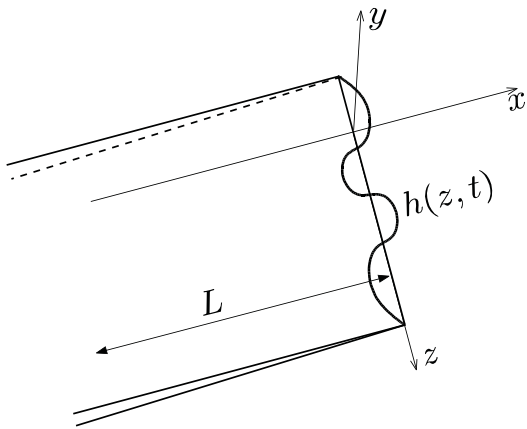}}
\caption{Schematic representation of a tensile crack in an infinite
body propagating in the plane $y=0$. The average penetration of the
crack front in the $x$-direction is $L$. The straight reference
front in the $z$-direction and the perturbation $h(z,t)$ around it
are also shown.} \label{fig:problem1}
\end{figure}

In order to write the equation of motion of a crack front, we first recall the form of its stress intensity factor~\cite{GaoRice89,PRE}. Consider a crack submitted to a tensile loading and propagating in the $(x,z)$ plane, with a front located at $x=h(z,t)$, $t$ being the time coordinate (see Fig.~\ref{fig:problem1}). The {\it static} stress intensity factor $K[h]$ was computed perturbatively to second order in $h$ and is given by~\cite{PRE}
\begin{equation}
\frac{K[h]}{K_0}=1-\frac{1}{8}h'^2(z)+\mathcal{PV}\int\frac{h(z')-h(z)}{(z'-z)^2}\left[1+\mathcal{PV}\int\frac{h(z'')-h(z')}{(z''-z')^2}\,\frac{\mathrm{d}z''}{2\pi}\right]\frac{\mathrm{d}z'}{2\pi}\,,
\end{equation}
which might be rewritten as
\begin{equation}
\frac{K[h]}{K_0}=1+\mathcal{H}[h]-\frac{1}{8}h'^2+\frac{1}{4}hh''+\mathcal{H}[h\mathcal{H}[h]]+ \mathcal{O}(h'^3)\,.
\label{eq:K}
\end{equation}
Here $K_0$ is the stress intensity factor for $h=0$, that is when the front is straight and parallel to the the $z$-axis, $h'=\partial h/\partial z$ is the spatial derivative of $h$ and $\mch$ is the Hilbert transform of the derivative,
\begin{equation}
\mathcal{H}[f](z)=\frac{1}{2\pi} \ \mathcal{PV} \! \int \frac{f'(z')}{z'-z}\dd{z'}=\frac{1}{2\pi} \ \mathcal{PV} \! \int \frac{f(z')-f(z)}{(z'-z)^2}\dd{z'}\,.
\end{equation}
Note that the perturbation expansion (\ref{eq:K}) of the stress intensity factor is invariant under translation of the front, $h\to h+\mathrm{cst}$. This expansion can be rewritten in the Fourier space as
\begin{equation}
\left(\frac{K[h]}{K_0}\right)_q=\delta(q) -\frac{1}{2}|q| h_q+\frac{1}{8}\int\left(2|q||\ell|-2\ell^2+\ell(q-\ell)\right)h_q h_{q-\ell}\frac{\mathrm{d}\ell}{2\pi}\,.
\end{equation}

\subsubsection{The equation of motion}

\begin{figure}[ht]
\centerline{\includegraphics[width=7cm]{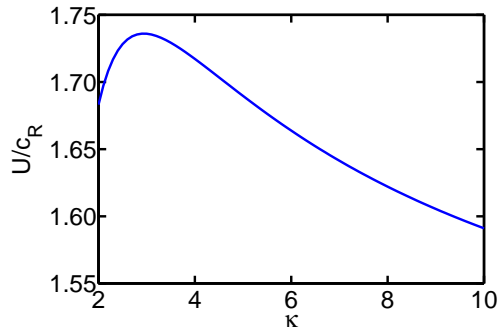}} \caption{The characteristic velocity $U(\kappa)$ in the equation of motion~(\ref{eq:motionfrac}), measured in units of the Rayleigh wave speed $c_\mathrm{R}$.}
 \label{fig:U*}
\end{figure}

The equation of motion of the crack front can be obtained by considering the dynamic energy release rate for a crack propagating at velocity slow compared to the characteristic wave speed of the material. According to the generalized Griffith energy criterion \cite{Freund}, the crack must grow in such a way that the energy release rate $G$ is equal to the dynamic fracture energy of the material $\Gamma(v)$. This material property may depend on the instantaneous crack tip speed \cite{Freund}; however, for low velocities, one can assume that the fracture energy is approximately constant $\Gamma(v)\simeq\Gamma$. In the case of a propagating tensile crack front, this reduces to~\cite{Freund}
\begin{equation}
G = \frac{1}{{2\mu }}{g  \left({v} \right)K  ^2 \left( {t,v = 0} \right)}  = \Gamma \, ,
\label{eq:Griffith}
\end{equation}
where $\mu$ is the Lam\'e shear coefficient and $v$ is the {\it instantaneous} local normal velocity of the crack front. The velocity dependent function $g(v)$, given in Appendix~A, is universal in the sense that it does not depend on the details of the applied loading, nor on the configuration of the body being analyzed; it depends only on the properties of the material through the dilatational ($c_\mathrm{d}$) and the shear ($c_\mathrm{s}$) wave speeds. Last, $K\left(t,v = 0\right)$ is the {\it static} stress intensity factor. A reformulation of Eq.~(\ref{eq:Griffith}) is the generalized Irwin criterion~\cite{Irwin},
\begin{equation}
\sqrt {g \left( {v} \right)} K\left[ h \right] = K_\mathrm{c} \, ,
\label{eq:Irwin}
\end{equation}
introducing the material toughness $K_\mathrm{c}\equiv\sqrt{2\mu \Gamma}$. At low velocities, one can expand
\begin{equation}
\sqrt {g (v)}  \simeq 1 - \alpha (\kappa)\frac{v}{c_\mathrm{R}} + \frac{1}{2}\alpha^2(\kappa)\left( \frac{v}{c_\mathrm{R}} \right)^2  + \mathcal{O}\left( \left(\frac{ v}{c_\mathrm{R}} \right)^3 \right) \, ,
\label{eq:g1}
\end{equation}
where $c_\mathrm{R}$ is the Rayleigh wave speed, $\alpha(\kappa)$ is a dimensionless function of the material parameter $\kappa =\left(c_\mathrm{d}/c_\mathrm{s}\right)^2$ (see Appendix~A). Eq.~(\ref{eq:Irwin}) now becomes
\begin{equation}
\frac{v}{c_\mathrm{R}} - \frac{1}{2}\alpha (\kappa)\left( \frac{v}{c_\mathrm{R}} \right)^2 \simeq \frac{1}{\alpha (\kappa)}\frac{K \left[ h \right] - K_\mathrm{c}}{K \left[ h\right]} \, ,
 \label{eq:Irwin2}
\end{equation}
leading to the $2^{nd}$ order expression for the velocity
\begin{equation}
\frac{v}{c_\mathrm{R}} \simeq \frac{1}{\alpha(\kappa)}\left[ {\frac{{K \left[
h \right] - K_\mathrm{c} }} {{K[h]}} +\frac{1}{2}\left(
\frac{{K \left[ h \right] - K_\mathrm{c} }} {{K[h]}}
\right)^2 } \right]
 \label{eq:vfrac} \, ,
\end{equation}
with a surprising simplification that makes the term inside the brackets independent of $\kappa$. This implies the equation of motion
\begin{equation}
\frac{1}{\sqrt {1 + h'^2}}\frac{\partial h}{\partial t} =U(\kappa) \left[\frac{K \left[ h \right] - K_\mathrm{c}}{K[h]} + \frac{1} {2}\left( \frac{K \left[ h \right]- K_\mathrm{c}}{K[h]} \right)^2 \right]H\left(K \left[ h \right] - K_\mathrm{c} \right) \qquad \textrm{[Fracture]} \, ,
 \label{eq:motionfrac}
\end{equation}
where the Heaviside function $H(\cdot)$ accounts for the irreversibility of crack opening, the term $ \sqrt{1 +h'^2 }$ is the ratio between the $x$-velocity and the normal velocity of the crack front, and $U(\kappa) = c_\mathrm{R}/ \alpha(\kappa)$ is a characteristic velocity. Note that $U(\kappa)$, plotted in Fig.~\ref{fig:U*}, is of the order of magnitude of the Rayleigh wave speed $c_\mathrm{R}$, which is very high. This gives a possible explanation for why even very close to the depinning threshold the movements of the interface are fast and therefore difficult to capture.

\subsection{Wetting}

\subsubsection{The contact angle}

\begin{figure}[ht]
\centerline{\includegraphics[width=7cm]{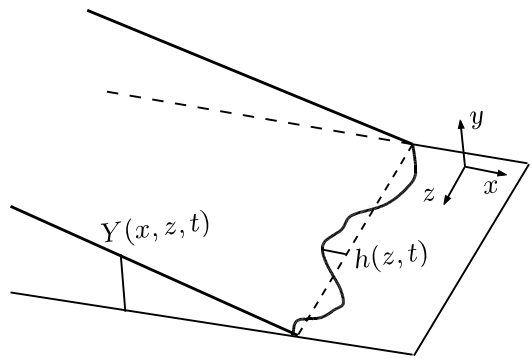}}
\caption{Schematic representation of a rough contact line. The straight reference line in the $z$-direction and the perturbation $h(z,t)$ around it are also shown.}
\label{fig:problem2}
\end{figure}

Consider a liquid film on a solid substrate, such that there is a triple contact line liquid/solid/gas where the thickness $Y(x,z,t)$ of the liquid layer vanishes. The position $x=h(z,t)$ gives the location of the contact line (Fig.~\ref{fig:problem2}). In order to establish its equation of motion, we first compute the contact angle as a function of $h$. We assume the motion of the contact line to be quasi-static. The length scales of interest are below the capillary length so that we neglect gravity. Moreover, the slope of the liquid layer is considered to be small. Within this framework, surface tension is solely involved in the shape of the layer and the thickness $Y(x,z,t)$ is a solution of the Laplace equation. The experimental device imposes the slope of the film $\theta_0$ far from the contact line. As a result, the problem is to solve
\begin{equation}
  \left\{\begin{array}{l}
      \displaystyle{\left(\frac{\partial^2}{\partial x^2}+\frac{\partial^2}{\partial z^2}\right)} Y=0\,,\smallskip\qquad x<h(z,t)\\
      Y(x=h(z,t),z)=0 \,,\smallskip\\
     \displaystyle{ \frac{\partial Y}{\partial x}(x=-\infty,z)=-\theta_0\, ,}
    \end{array}\right.
\end{equation}
and more precisely to find the value of the contact angle $\theta=-{\partial Y}/{\partial n}$ at the location $(x=h(z,t),z)$.

This problem is the same as the one solved in~\cite{Sultan1,Sultan2} in the context of the diffusion-limited evaporation of a thin film. Translating into the terms of wetting contact line problem, this gives
\begin{equation}
  \frac{\theta[h]}{\theta_0}=1+2\mathcal{H}[h]+\frac{1}{2}h'^2+hh''+4\mathcal{H}[h\mathcal{H}[h]]
-h'^2\mathcal{H}[h]+\left(h^2 \mathcal{H}h\right)''+ \mathcal{H}\left(h^2h''\right)+8\mathcal{H}[h\mathcal{H}[h\mathcal{H}[h]]]
+ \mathcal{O}(h'^4)\, .
\label{theta}
\end{equation}
At first and second order in $h$, the contact angle has the same terms as the stress intensity factor but with different numerical prefactors. This form was previously derived at second order in~\cite{Golestanian}. Note that the derivation of this expression in~\cite{Sultan1,Sultan2} suggests that higher order terms are polynomial in $h'^2$ and in the operator $\mch[h' \mathbf{\cdot}]$. As will be discussed in the conclusion, this remark has implications on the relevance of high order terms in the equation of motion of the contact line. Eventually, for completeness we give the Fourier transform of the contact angle,
\begin{eqnarray}
 \left(\frac{\theta[h]}{\theta_0}\right)_q&=&\delta(q)-|q|h_q+\int  \left\{\frac{1}{2}\ell (q-\ell)+|q||\ell|-\ell^2\right\}h_\ell h_{q-\ell}\frac{\mathrm{d\ell}}{2\pi}\nonumber\\
 &+&\int \left\{-\frac{1}{2}|m|\ell (q-m-\ell)+\frac{1}{2}q^2|m|+\frac{1}{2}|q|m^2-|q||m||m+\ell|\right\}h_\ell h_m h_{q-m-\ell}\frac{\mathrm{d}\ell}{2\pi}\frac{\mathrm{d}m}{2\pi}\,.
\end{eqnarray}

\subsubsection{Equations of motion}

According to the underlying microscopic physics, various equations of motion were proposed. In the model by Blake~\cite{blake}, the normal velocity derives from the surface energy of the system
\begin{equation}
\frac{\partial h}{\partial t}=U \sqrt{1+h'^2}\left(\theta[h]^2-\theta_\mathrm{eq}^2\right)H(\theta[h]-\theta_\mathrm{eq})\, ,
\quad \textrm{[Blake]}
 \label{eq:B}
\end{equation}
where $U$ is the characteristic hopping velocity of the liquid molecules at the contact-line. The term $ \sqrt{1 +h'^2 }$ is again the ratio between the $x$-velocity and the normal velocity of the line. $\theta_\mathrm{eq}$ is the equilibrium contact angle. The Heaviside $H$ function might be used to account for the hysteresis in contact angle; $\theta_\mathrm{eq}$ then stands for the advancing contact angle.

In de Gennes~\cite{degennes86}, the viscous dissipation is balanced by the the gain in surface energy
\begin{equation}
\frac{\partial h}{\partial t}=U \sqrt{1+h'^2}  \, \theta [h] \left(\theta[h]^2-\theta_\mathrm{eq}^2\right)H(\theta[h]-\theta_\mathrm{eq}) \, .
\quad \textrm{[de Gennes]}
 \label{eq:JdG/PV}
\end{equation}
The capillary velocity $U=\gamma/\mu$ is the ratio of the surface tension $\gamma$  of the liquid and its viscosity $\mu$.

In the purely hydrodynamical approach due to Cox-Voinov~\cite{Cox,Voinov},
\begin{equation}
\frac{\partial h}{\partial t}=U \sqrt{1+h'^2} \left(\theta[h]^3-\theta_\mathrm{eq}^3\right)H(\theta[h]-\theta_\mathrm{eq}) \, ,
\textrm{\quad [Cox-Voinov]}
 \label{eq:CV}
\end{equation}
where $U$ is again the capillary velocity. Note that the model with a diffuse interface proposed by Pomeau and Pismen~\cite{pismen,pomeaurev} yields a similar macroscopic form.

\subsection{The annealed noise approximation}

In the case of propagation of a crack front through a heterogeneous material or wetting of a disordered substrate, the standard way to introduce the effect of heterogeneity in the equations of motion consists of modeling the material toughness $K_\mathrm{c}(z,h)$ or the equilibrium contact angle $\theta_\mathrm{eq}(z,h)$ as random functions which are separated into a constant and a fluctuating part as
\begin{eqnarray}
K_\mathrm{c}(z,h) &=& K^*  + K^* \hat{\eta} \left( {z,h} \right)\,,\\
\theta_\mathrm{eq}(z,h) &=& \theta^*  + \theta^* \hat{\eta} \left( {z,h} \right)\,,
\end{eqnarray}
where $\hat\eta(z,h)$ is a random noise with a zero mean and is assumed to be short-range correlated.

Our approach is based on the assumption that one can approximate the noise term for the moving front, where $h \simeq Vt$, by $\hat{\eta}(x,h)\simeq \hat{\eta} (x,Vt) = \eta (x,t)$. This approximation amounts to saying that the fluctuations in the front velocity are small compared to the imposed driving velocity $V$. Also, we do keep nonlinear terms, since we claim (and will justify later) that they play an important role in roughening the interface. Obviously a linear equation of the kind described above, i.e. taking into consideration only the linear term in $K[h]$ (or in $\theta[h]$), would not yield any roughness, and actually even if the KPZ nonlinearities (i.e. $h'^2$ terms) are kept, we would also end up with a smooth surface, or at most logarithmically rough. This is a special case of the so called Fractal KPZ equation studied previously in \cite{FKPZ03}.

The different equations of motion (\ref{eq:motionfrac},\ref{eq:B},\ref{eq:JdG/PV},\ref{eq:CV}), for fracture and wetting respectively, have the same mathematical structure, so that we can study both systems simultaneously. To this purpose, we rescale for fracture $t\rightarrow t/U$ and for wetting $h \rightarrow h/4$ and $t\rightarrow t/(4U)$. When inserting the form $K[h]$ (or $\theta[h]$) in each equation, expanding to $2^{nd}$ order in $h$ and assuming the noise amplitude to be of the same order as $h^2$, all the equations can be recast in the general form

\begin{equation}
\frac{\partial h}{\partial t}=v+\mathcal{H}[h]-\frac{a}{4}
h'^2+\frac{1}{4}hh''+\mathcal{H}[h\mathcal{H}[h]]+\frac{b}{4}\left(\mathcal{H}[h]\right)^2  +  \eta \left({z,t} \right) \,,
 \label{eq:motion}
\end{equation}
where the dimensionless velocity of the unperturbed front $v$ and the coefficients $a$ and $b$ are given by
\begin{eqnarray}
a= - \frac{3K^{*2} - 10K_0 K^* + 6K_0^2}{2K^*\left({2K_0 - K^*}\right)} &,& \quad b= - 2\frac{4K_0 - 3K^*}{2K_0 - K^*} \,, \quad v= \frac{{\left( {K_0  - K^* } \right)\left( {3K_0 - K^* } \right)}}{{2K^* \left( {2K_0  - K^* } \right)}}\,,\textrm{\quad [Fracture]}\\
a=-\frac{3\theta_0^2-\theta^{*2}}{8\theta_0^2} &,& \quad b=1\, , \quad v=\frac{\theta_0^2-\theta^{*2}}{\theta_0^2}\,,\textrm{\quad [Blake]}\\
a=-\frac{2\theta_0^2-\theta^{*2}}{3\theta_0^2-\theta^{*2}} &,& \quad b=\frac{12\theta_0^2}{3\theta_0^2-\theta^{*2}}\,, \quad v=\frac{\theta_0^2-\theta^{*2}_\mathrm{eq}}{3\theta_0^2-\theta^{*2}}\,,\textrm{\quad [de Gennes]}\\
a=-\frac{4\theta_0^3-\theta^{*3}}{6\theta_0^3} &,& \quad b=4\,, \quad v=\frac{\theta_0^3-\theta^{*3}}{3\theta_0^3}\,,\textrm{\quad [Cox-Voinov]}
\end{eqnarray}
Resulting from Equations~(\ref{eq:motionfrac},\ref{eq:B},\ref{eq:JdG/PV},\ref{eq:CV}) respectively. A comparison between these different cases is given below in Fig.~\ref{fig:abv}. Notice that there is a real difference between the prefactors in the different cases in terms of sign, existence of zero-crossing, and general behavior as function of $K^*/K_0$ (or $\theta^*/\theta_0$).
\begin{figure}[ht]
\centerline{\includegraphics[width=6cm]{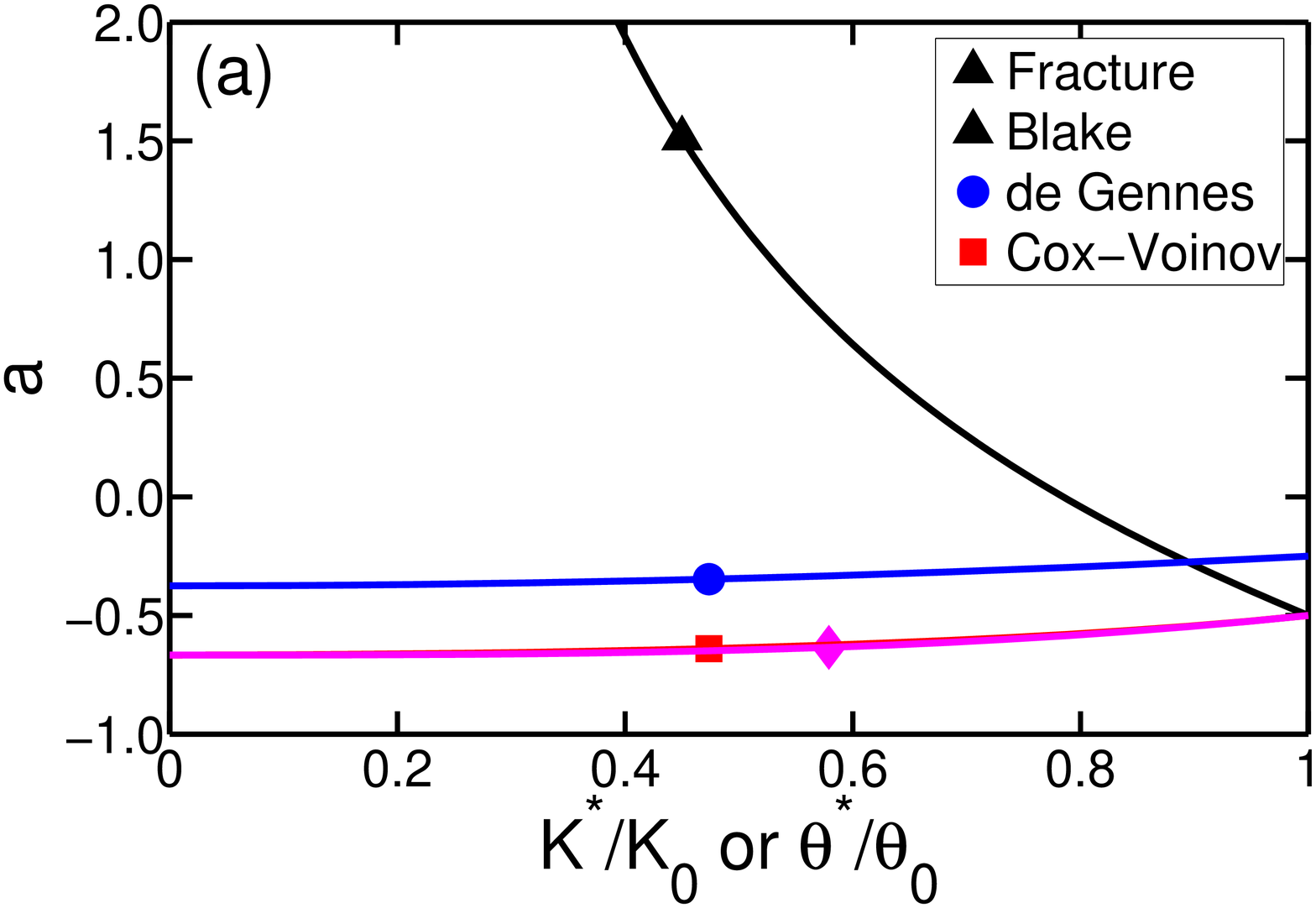}
\includegraphics[width=6cm]{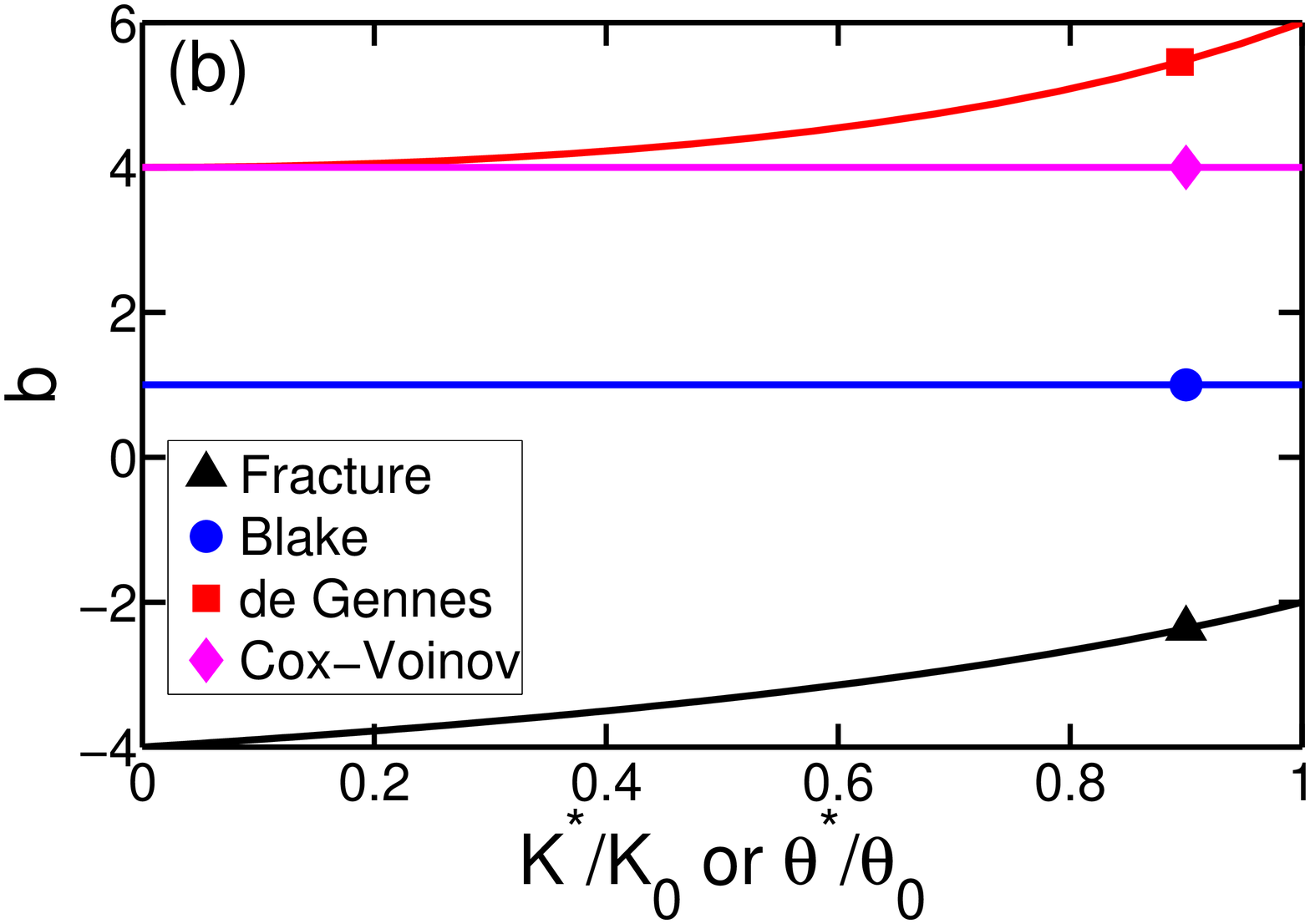}
\includegraphics[width=6cm]{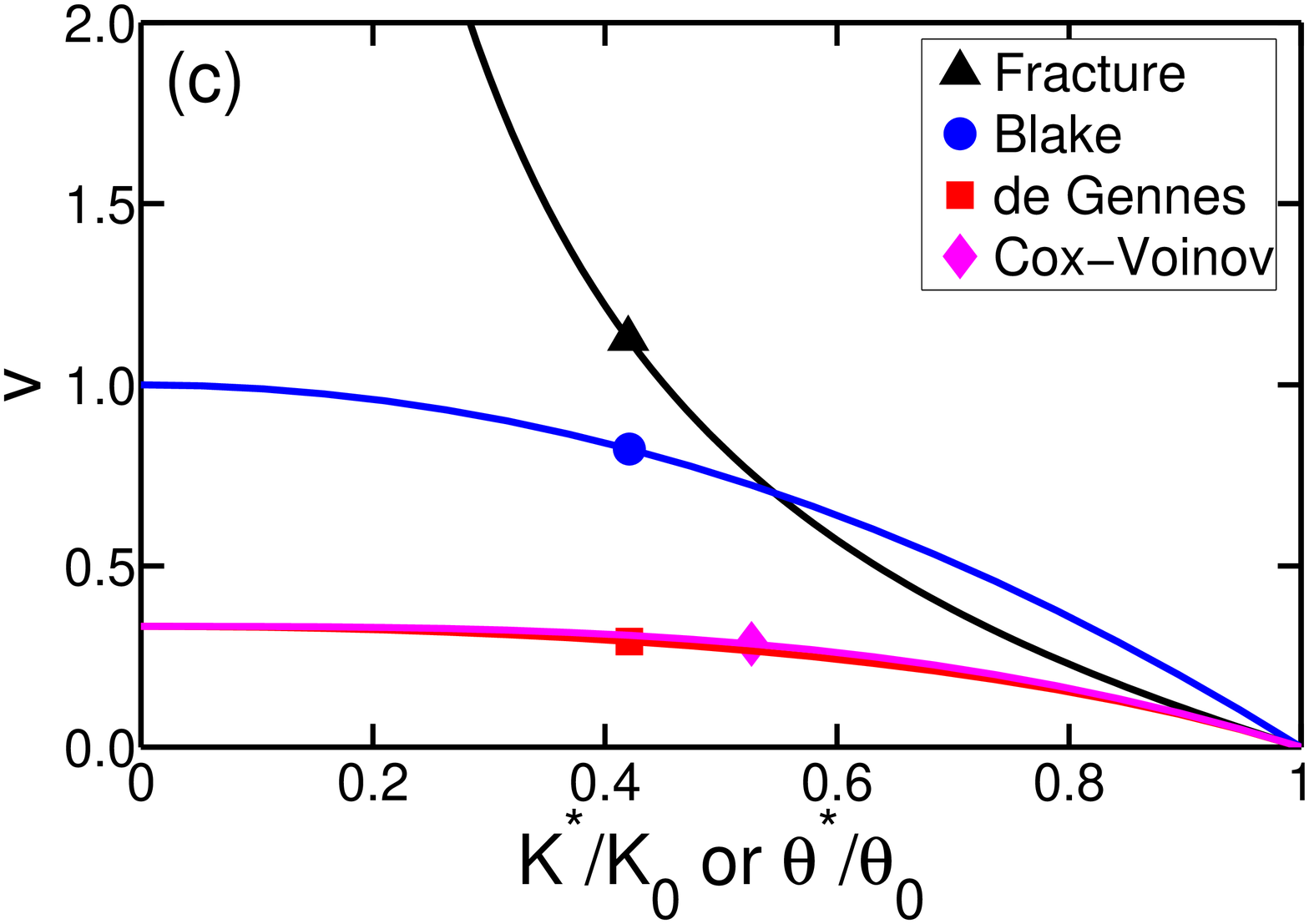}}
\caption{The prefactors of the nonlinear terms in the equation of motion (\ref{eq:motion}) as functions of the ratio of the average toughness to the driving stress intensity factor $K^*/K_0$ (or the ratio of the average equilibrium contact angle to the driving angle $\theta_\mathrm{eq}^*/\theta_0$ for wetting).}
\label{fig:abv}
\end{figure}

Interestingly, the velocity of the unperturbed front $v$ in Eq.~(\ref{eq:motion}) can be eliminated, by transforming into a co-moving coordinate system, i.e. $h \rightarrow h+vt$, and so in the following we omit it. Finally, the noise correlations are described by
\begin{equation}
\left\langle \eta \left( z,t \right)\eta \left( z',t' \right) \right\rangle  = 2D\delta \left( z - z' \right)\delta \left( t - t' \right)\,,
\end{equation}
where $D$ is the variance of the noise.

\section{The Self Consistent Expansion}

\subsection{Description of the method}

The SCE method is based on going over from the Fourier transform of the equation in Langevin form to a Fokker-Planck form and on constructing a self-consistent expansion of the distribution of the field concerned. We thus consider the equation of motion~(\ref{eq:motion}) (transformed in the co-moving frame) in Fourier components
\begin{equation}
\frac{\partial h_q }{\partial t} =  - c_q h_q  -\sum\limits_{\ell ,m} {M_{q\ell m} h_\ell  h_m }  + \eta _q\left( t \right)\,,
\label{eq:motionFourier}
\end{equation}
where $c_q = \left|q\right|$ and
\begin{equation}
M_{q\ell m} = -\frac{1}{4\sqrt \Omega} \left[\left| q \right|\left| \ell  \right| -  a\,\ell m  -b\left| \ell  \right|\left| m \right| -\ell^2 \right]\delta_{q,\ell+ m} \, ,
\label{eq:M}
\end{equation}
$\Omega$ being the linear length of the front. Note that in contrast to the KPZ problem $M_{q\ell m}$ has only the symmetry $M_{q\ell m} = M_{-q,-\ell,-m}$. Here, $\eta_q(t)$ is a noise term with zero average described by its variance
\begin{equation}
\left\langle {\eta _q \left( t \right)\eta _{q'} \left({t'} \right)} \right\rangle  = 2D\delta\left(q+q'\right) \delta \left( {t -t'} \right)\,.
\end{equation}
Rewriting Eq.~(\ref{eq:motionFourier}) in a Fokker-Planck form, one gets
\begin{equation}
\frac{\partial P}{\partial t} + \sum\limits_q \frac{\partial}{\partial h_q}\left[D \frac{\partial}{\partial h_{ - q}} +
c_q h_q  + \sum\limits_{\ell,m} M_{q\ell m} h_\ell  h_m\right]P  = 0\,, \label{eq:motionFokker}
\end{equation}
where $P(\{h_q\},t)$ is the probability functional for having a height configuration $\{h_q\}$ at time $t$.

The expansion is formulated in terms of the steady-state structure factor (also called the two-point function)
\begin{equation}
\phi_q = \left\langle {h_{-q}h_q}\right\rangle\,,
\end{equation}
and its corresponding steady-state decay rate $\omega_q$, which describes the rate of decay of a disturbance of wave vector $q$ in steady state and is defined by
\begin{equation}
\omega_q^{-1} = \frac{\int_0^\infty \left\langle {h_{-q}(0) h_q(t)}\right\rangle dt}  {\left\langle {h_{-q} h_q }\right\rangle}
\label{eq:omega} \, .
\end{equation}
From the scaling form (\ref{scaling}), it follows that for small $q$'s, $\phi_q$ and $\omega_q$ behave as power laws in $q$, namely
\begin{eqnarray}
\phi_q &=& A|q|^{-\Gamma}\,,
\label{eq:phiq}\\
\omega_q &=& B|q|^z\,,
\label{eq:omegaq}
\end{eqnarray}
where $z$ is the dynamic exponent, and $\Gamma$ is related to the roughness exponent by $\zeta=(\Gamma-1)/2$ \cite{Schmittbuhl95a}.

The main idea of SCE is to write the Fokker-Planck equation $\partial P/\partial t= OP$ in the form
\begin{equation}
\frac{\partial P}{\partial t}= \left[ O^{(0)} + O^{(1)} + O^{(2)} \right]P \, ,
\end{equation}
where $O_0$, $O_1$ and $O_2$ are zero, first and second order operators in some parameter. The evolution operator $O_0$ is chosen to have a simple form
\begin{equation}
O_0  =  -\sum\limits_q {\frac{\partial }{{\partial h_q }}\left( {\tilde \omega_q \tilde \phi_q
\frac{\partial }{{\partial h_{ - q} }} + \tilde \omega_q h_q } \right)}
\label{eq:O0}\, ,
\end{equation}
whose corresponding zeroth order solution for $P^{(0)}$ is a Gaussian given by
\begin{equation}
P^{(0)} \left(\{ h_q \}\right) \propto \exp{\left[ -\frac{1}{2} \sum_p \frac{h_{-p} h_p} {\tilde \phi_p} \right]} \, ,
\end{equation}
implying that at leading order the two point function is given by $\langle h_{-q}h_q \rangle ^{(0)} = \tilde \phi_q$. Note however that $\tilde \phi_q$ and $\tilde \omega_q$ are still unknown functions.
Next, an equation for the two-point function is obtained. The expansion has the form $\phi_q = \tilde \phi_q + e_q\left\{\tilde\phi_p,\tilde\omega_p\right\}$, where $e_q$ is a functional of all $\tilde \phi$'s and $\tilde \omega$'s. This reflects the fact that the lowest order in the expansion is exactly the unknown $\phi_q$. In the same way, an expansion for $\omega_q$ is given by $\omega_q = \tilde \omega_q + d_q\left\{\tilde \phi_p ,\tilde \omega_p\right\}$. The two-point function and the characteristic frequency are determined by the requirement that the corrections to the zero order vanish, and thus given by the two coupled equations
\begin{eqnarray}
e_q \left\{ \tilde \phi_p ,\tilde \omega_p \right\} &=& 0
\label{e} \, ,\\
d_q \left\{ \tilde \phi_p ,\tilde \omega_p \right\} &= &0
\label{d} \, .
\end{eqnarray}
Eqs.~(\ref{e},\ref{d}) can be solved exactly in the asymptotic limit of small $q$'s yielding the required scaling exponents governing the steady-state behavior and the time evolution.

\subsection{Equations for the structure factors}

Since the symmetries of the kernel $M_{q \ell m}$ as defined in Eq.~(\ref{eq:M}) are different from those of the KPZ equation, we cannot use the SCE equations obtained in \cite{SCE} and we need to derive them for the present problem. We first rewrite the Fokker-Planck equation (\ref{eq:motionFokker}) along the lines presented above as
\begin{equation}
\frac{\partial P}{\partial t} + \sum\limits_p \frac{\partial }{\partial h_p }\left[ \tilde \omega_p \tilde \phi_p \frac{\partial }{\partial h_{-p}} + \tilde \omega_p h_p  + \sum\limits_{\ell,m} M_{p\ell m} h_{\ell} h_m + (D  - \tilde \omega_p \tilde \phi_p)\frac{\partial }{\partial h_{-p}} + (c_p  - \tilde \omega_p) h_p \right] P = 0
\label{eq:SCE1} \, ,
\end{equation}
where $\tilde \phi_p$ and $\tilde \omega_p$ are unknown functions. The $M$ term in Eq.~(\ref{eq:SCE1}) is considered as the operator $O^{(1)}$ and the differences $(D-\tilde \omega_p \tilde \phi_p)$ and $(c_p-\tilde \omega_p)$ are considered
as $O^{(2)}$. In a similar way we can expand the probability functional
\begin{equation}
 P(\{h_q\},t) = P^{(0)} + P^{(1)} + P^{(2)} + \cdots \, ,
\end{equation}
from which follows an expansion for an expectation value for an arbitrary observable $\mathcal{F}$ to a prescribed order $n$
\begin{equation}
 \langle \mathcal{F} \rangle ^{(n)} \equiv \int \mathcal{F}(\{h_q\}) \sum_{m=0}^{n} P^{(m)}(\{h_q\}) \mathcal{D}h_q \, .
\end{equation}
In order to calculate such averages we multiply Eq.~(\ref{eq:SCE1}) by $\mathcal{F}(\{h_q\})$ and integrate with respect to $h_q$'s. After some integrations by parts we get
\begin{eqnarray}
 \frac{{\partial \left\langle \mathcal{F} \right\rangle ^{\left( n \right)} }}{\partial t} + \sum\limits_p {\left[ {\tilde \omega_p \tilde \phi_p  \left\langle {\frac{{\partial ^2 \mathcal{F}}}{{\partial h_{-p} \partial h_p }}} \right\rangle ^{(n)} - \tilde \omega_p \left\langle {h_p \frac{{\partial \mathcal{F}}}{{\partial h_p }}} \right\rangle ^{(n)} - \sum\limits_{\ell,m} {M_{p\ell m} \left\langle {h_{\ell} h_m \frac{{\partial \mathcal{F}}}{{\partial h_p }}} \right\rangle ^{(n - 1)} } } \right.}  \nonumber \\
 \left. { + \left( {D - \tilde \omega_p \tilde \phi_p } \right)\left\langle {\frac{{\partial ^2 \mathcal{F}}}{{\partial h_{-p} \partial h_p }}} \right\rangle ^{(n-2)} - (c_p  - \tilde \omega _p) \left\langle {h_p \frac{{\partial \mathcal{F}}}{{\partial h_p }}} \right\rangle ^{(n - 2)} } \right] = 0
\label{eq:F} \, ,
\end{eqnarray}
where we have added the superscript $(n)$ to keep track of the order of the quantities with respect to the perturbative structure. At first we are interested in steady state properties and therefore drop any time dependence from the last equation. Later on, when deriving the equation for the characteristic decay rate the time dependence will be retrieved. Then, being interested in the $2$-point function, we insert
$\mathcal{F}=\frac{1}{2}h_q h_{-q}$ into Eq.~(\ref{eq:F}). To zeroth order we get
\begin{equation}
\tilde \omega_q \tilde \phi_q  - \tilde \omega_q \langle{h_{-q} h_q }\rangle^{(0)}=0 \, ,
\end{equation}
from which follows $\langle{h_{-q} h_q }\rangle^{(0)} = \tilde \phi_q$. Similarly, considering $\mathcal{F}=\frac{1}{2}h_q h_{-q}$ to first order gives
\begin{equation}
\left\langle {h_{-q} h_q } \right\rangle ^{(1)} =
\tilde \phi_q - \frac{1}{\tilde \omega_q} \sum\limits_{\ell,m} M_{q\ell m} \langle h_\ell h_m h_{-q} \rangle ^{(0)} \, .
\end{equation}
In order to calculate the correction we need to know the average of the $3$-point function evaluated at zeroth order. To achieve this we insert $\mathcal{F}=h_L h_M h_{-q}$ (with $L+M=q$) into Eq.~(\ref{eq:F}) and consider zeroth order contributions. This gives exactly zero, and so $\langle{h_{-q} h_q }\rangle^{(1)}=\langle{h_{-q} h_q }\rangle^{(0)}$. Thus, in order to get subleading corrections to the $2$-point function wee need to go to second order
\begin{equation}
\tilde \omega _q \left\langle {h_{-q} h_q} \right\rangle^{(2)} = \tilde \omega _q
\left\langle {h_{-q} h_q} \right\rangle^{(0)}  + D  - c_q
\left\langle {h_{-q} h_q} \right\rangle ^{(0)} - \sum\limits_{\ell,m}
{M_{q\ell m} \left\langle{h_\ell h_m h_{-q}}\right\rangle ^{(1)}}.
\label{2P2}
\end{equation}
As above, in order to calculate $\left\langle{h_\ell h_m h_{-q}}\right\rangle^{(1)}$ we insert $\mathcal{F}=h_L h_M h_{-q}$ (with $L+M=q$) into Eq.~(\ref{eq:F}) and consider first order contributions. This gives
\begin{equation}
\left\langle {h_\ell h_m h_{- q}} \right\rangle ^{(1)}  =  - \tilde \phi_q
\frac{\tilde \phi_m \left(M_{\ell,-m,q}  + M_{\ell,q,-m} \right) + \tilde \phi_\ell \left(M_{m,-\ell,q} + M_{m,q,-\ell} \right)}{\tilde \omega_\ell + \tilde \omega_m  + \tilde \omega_q}
- \frac{\tilde \phi_\ell \tilde \phi _m \left(M_{-q,-\ell,-m} + M_{-q,-m,-\ell}\right)} {\tilde \omega_\ell + \tilde \omega_m  + \tilde \omega_q}\,. \label{3P1}
\end{equation}
Plugging this result into Eq.~(\ref{2P2}) and using the symmetry of $M_{q \ell m}$ gives
\begin{eqnarray}
 \tilde \omega _q \left\langle h_{-q} h_q \right\rangle ^{(2)} &=& \tilde \omega _q \left\langle h_{-q} h_q\right\rangle ^{(0)} + D
- c_q \left\langle h_{-q} h_q \right\rangle ^{(0)} + \sum\limits_{\ell,m} \frac{M_{q\ell m} \left(M_{q\ell m}  + M_{qm\ell} \right)\tilde \phi_\ell \tilde \phi_m}{\tilde \omega_\ell  + \tilde \omega_m  + \tilde \omega_q}  \nonumber \\
 &+&\phi_q \sum\limits_{\ell ,m}\frac{\left(M_{q\ell m}+M_{qm\ell}\right)\left(M_{\ell, - m,q}  + M_{\ell ,q, - m}\right)\phi _m} {\omega_\ell + \omega_m + \omega_q }.
\end{eqnarray}
The last step in the Self-Consistent Expansion is to impose self-consistency on this second order expression. This means choosing $\tilde \phi_q$ such that the two-point function to second order would be exactly equal to the zeroth order result. This is achieved by identifying $\left\langle h_{-q}h_q \right\rangle^{(2)} = \left\langle h_{-q}h_q \right\rangle ^{(0)} = \phi_q$ and so the rest, denoted $e_q\{\phi_p , \omega_p\}$ in Eq.~(\ref{e}), is set to zero~:
\begin{equation}
 D  - c_q \phi_q +
\phi_q \sum\limits_{\ell ,m}\frac{\left(M_{q\ell m}+M_{qm\ell}\right)\left(M_{m, - \ell,q}  + M_{m ,q, - \ell}\right)\phi _\ell} {\omega_\ell + \omega_m + \omega_q }
+ \sum\limits_{\ell ,m} \frac{M_{q\ell m} \left(M_{q\ell m}  + M_{qm\ell }\right)\phi _\ell  \phi _m }{\omega_\ell + \omega_m  + \omega_q}= 0.
\end{equation}
In a similar way we can derive an equation for $\omega_q$ corresponding to Eq.~(\ref{d}). Since $\omega_q$ is related to dynamical properties (see Eq.~(\ref{eq:omega})) we need to consider time-dependent quantities. Therefore, we can plug $\mathcal{F}=h_{-q}(0) h_q (t)$ into Eq.~(\ref{eq:F}) and proceed as above. As shown in \cite{SCE,TCN04}, an alternative way is to use Herring's consistency equation \cite{Herring} which reads
\begin{equation}
c_q  - \omega _q  + \sum\limits_{\ell ,m}\frac{\left(M_{q\ell m}+M_{qm\ell}\right)\left(M_{m, - \ell,q}  + M_{m ,q, - \ell}\right)\phi _\ell} {\omega_\ell + \omega_m } = 0.
\end{equation}

The last two equations allow us to determine the two unknown function $\phi_q$ and $\omega_q$. But before proceeding it is useful to rewrite them in the following way
\begin{eqnarray}
D        &-& |q| \phi_q + I_1(q)\phi_q + I_2(q) = 0 \, ,\label{eq:Static2} \\
\omega_q &-& |q| + J(q) = 0 \,,\label{eq:Herring2}
\end{eqnarray}
with
\begin{eqnarray}
 I_1(q) &=& \frac{1}{32\pi} \int{d\ell {\textstyle{\frac{\left[ {|q| \left(\left| \ell  \right| + \left| {q - \ell } \right|
\right) - 2a\ell \left(q-\ell\right) + 2b\left| \ell \right|\left|
{q - \ell } \right| - \ell ^2  - \left(q-\ell\right)^2 }
\right]\left[ {\left| {q - \ell } \right|\left( {\left| \ell \right|
+ \left| q \right|} \right) + 2a\ell q + 2bq\left| \ell \right| -
\ell ^2 - q^2} \right]}{\omega _q  + \omega_\ell + \omega _{q-\ell } }}}\phi_\ell} \, , \label{eq:integrals1a} \\
 I_2(q) &=& \frac{1}{16\pi} \int{d\ell {\textstyle{ \frac{\left[ {\left|q
\right|\left( {\left| \ell  \right| + \left| {q - \ell } \right|}
\right) - 2a\ell \left( {q - \ell } \right) + 2b\left| \ell
\right|\left| {q - \ell } \right| - \ell ^2  - \left( {q - \ell }
\right)^2 } \right]^2}{\omega _q  + \omega _\ell   + \omega _{q
- \ell } }}}\phi _\ell  \phi _{q - \ell } } \, ,\label{eq:integrals1b}\\
 J(q) &=&  \frac{1}{32\pi} \int {d\ell {\textstyle{\frac{\left[ {\left| q \right|\left( {\left| \ell
\right| + \left| {q - \ell } \right|} \right) - 2a\ell \left( {q -
\ell } \right) + 2b\left| \ell  \right|\left| {q - \ell } \right| -
\ell ^2  - \left( {q - \ell } \right)^2 } \right]\left[ {\left| {q -
\ell } \right|\left( {\left| \ell  \right| + \left| q \right|}
\right) + 2a\ell q + 2bq\left| \ell  \right| - \ell ^2  - q^2 }
\right]}{\omega _\ell   + \omega _{q - \ell } }}}\phi _\ell}
\, . \label{eq:integrals1c}
\end{eqnarray}
Finally, it is interesting to mention here that
Eq.~(\ref{eq:Static2}) can be understood as emanating from the short
time balance of the original equation, while Eq.~(\ref{eq:Herring2})
comes from its long time balance \cite{TCN04}.

\subsection{General solutions for the scaling exponents}

Eqs.~(\ref{eq:Static2},\ref{eq:Herring2}) can be solved exactly in the asymptotic limit (i.e. for small $q$'s) to yield the required scaling exponents governing the steady-state behavior and the time evolution. The difficulty here arises from the fact that the integrals involved, $I_1(q)$, $I_2(q)$, and $J(q)$, have contributions from large $\ell$'s as well as from small $\ell$'s. Therefore, one must consider the contribution of the large $\ell$ integration domain on the small $q$ behavior of the integrals
(\ref{eq:integrals1a}-\ref{eq:integrals1c}). For this, we break up the integrals $I_i(q)$ and $J(q)$ into the sum of two contributions $I_i^<(q)+I_i^>(q)$, and $J^<(q)+J^>(q)$, corresponding to domains of integration over $\ell<\Lambda$ or $\ell>\Lambda$ respectively. We expand $I_i^>(q)$ and $J^>(q)$ for small $q$'s and obtain the leading small $q$  behavior of the integrals
\begin{eqnarray}
  I_1^> (q) &=& A_1 \left| q \right| - B_1 \left| q \right|\omega _q  + C_1 q^2  \, ,
  \label{eq:integrals21}\\
  I_2^> (q) &=& A_2 + B_2 \left| q \right| + C_2 q^2  - D_2 \left| q \right|\omega_q  \, ,
  \label{eq:integrals22}\\
  J^> (q)   &=& A_3 \left| q \right| - B_3 \left| q \right|\omega _q  + C_3 q^2\, ,
\label{eq:integrals23}
\end{eqnarray}
where the leading order coefficients $A_1,A_2,A_3$ are
\begin{eqnarray}
  A_1 = A_3 &=& \frac{(a + b - 1)(2b + 1)}{16\pi} \int_\Lambda ^\infty {\ell^3 \frac {\phi_\ell}{\omega_\ell} d\ell } \, ,
  \label{eq:A1}\\
  A_2 &=& \frac{(a + b - 1)^2}{16\pi} \int_\Lambda ^\infty {\ell ^4 \frac{\phi_\ell ^2}{\omega_\ell}d\ell } \, ,
  \label{eq:A2}
\end{eqnarray}
and so their exact values generally depend on the cutoff $\Lambda$. Also, note that $B_3=0$ (it was kept for clarity). Using these results, Eqs.~(\ref{eq:Static2},\ref{eq:Herring2}) reduce to
\begin{eqnarray}
D &+& A_2  - \left( {1 - A_1 } \right)\left| q
\right|\phi _q  + I_1^ <  \left( q \right)\phi _q  + I_2^ <  \left( q \right) = 0\,, \label{eq:Static3} \\
\omega _q &-& \left( {1 - A_1 } \right)\left| q \right| + J^ <
\left( q \right) = 0 \,.\label{eq:Herring3}
\end{eqnarray}
The advantage of Eqs.~(\ref{eq:Static3},\ref{eq:Herring3}) over Eqs.~(\ref{eq:Static2},\ref{eq:Herring2}) is that at the mere price of renormalizing some constants in both equations, we are left with the integrals $I_1^<(q)$, $I_2^<(q)$ and $J^<(q)$
that can be calculated explicitly for small $q$'s since the power-law form for $\phi_\ell$ and $\omega _\ell$ for small $\ell$'s (\ref{eq:phiq},\ref{eq:omegaq}) can be used. The treatment of Eqs.~(\ref{eq:Static3},\ref{eq:Herring3}) is carried on by studying the various possibilities of balancing the dominant order for small $q$. Note also that the small $q$-dependence of each of the integrals $I_i^<(q)$ and $J^<(q)$ depends on the convergence of the integrals without cutoffs. So, to leading order in $q$ one gets
\begin{eqnarray}
I_1^< (q) &\sim& \left\{
\begin{array}{ll}
 E_1|q| & \quad 4-\Gamma-z > 0 \\
 F_1|q|^{5-\Gamma-z} &\quad 4-\Gamma-z < 0
 \end{array} \right.\,,
\label{eq:integrals3}\\
I_2^<  \left( q \right) &\sim& \left\{ \begin{array}{ll}
 E_2 &\quad 5-2\Gamma-z > 0 \\
 F_2|q|^{5-2\Gamma-z} &\quad 5-2\Gamma-z < 0
 \end{array} \right.\,,
\label{eq:integrals4}\\
 J^<  \left( q \right) &\sim& \left\{
\begin{array}{ll}
 E_1|q| &\quad 4-\Gamma-z > 0 \\
 F_3|q|^{5-\Gamma-z}  &\quad 4-\Gamma-z < 0
 \end{array} \right.\,,
\label{eq:integrals5}
\end{eqnarray}

\begin{figure}[ht]
\centerline{\includegraphics[width=7cm]{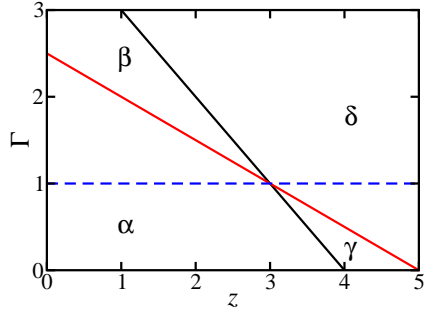}}
\caption{The possible domains for the exponents. Definition of the different sectors corresponding to the division of the $(\Gamma,z)$-plane by the curves $4-\Gamma-z=0$ and $5-2\Gamma-z=0$, with $\Gamma>0$ and $z>0$. Note that the physical domain of the roughness of the front is given by $1\le \Gamma< 3$, and thus solutions in sector $\gamma$ are not allowed.}
\label{fig:sectors}
\end{figure}

Let us consider the quadrant of the $(\Gamma,z)$-plane defined by $\Gamma>0$ and $z>0$, where solutions may be expected. The lines $4-\Gamma-z=0$ and $5-2\Gamma-z=0$ divide this quadrant into four sectors (see Fig.~\ref{fig:sectors}). The classical method \cite{SCE} is to investigate each sector separately and to decide whether or not a solution might exist there. In sector $\alpha$ defined by $\Gamma<(5-z)/2$ and $\Gamma<(4-z)$, one can rewrite Eqs.~(\ref{eq:Static3},\ref{eq:Herring3}) as
\begin{eqnarray}
 D + A_2 + E_2 - A\left( {1 - A_1 - E_1} \right) |q|^{1-\Gamma} &=& 0
\label{eq:Static4}\,,\\
 B|q|^z - \left( {1 - A_1  - E_1} \right)|q| &=& 0
\label{eq:Herring4}\,.
\end{eqnarray}
Eqs.~(\ref{eq:Static4},\ref{eq:Herring4}) are self-consistent in the small $q$ limit only if $\Gamma=z=1$. These values are allowed by the defining conditions of this sector, thus they are possible solutions. In sector $\beta$ defined by $(5-z)/2<\Gamma<(4-z)$, the difference from sector $\alpha$ lies in Eq.~(\ref{eq:Static4}) which is now rewritten as
\begin{equation}
 D + A_2 - A\left( {1 - A_1 - E_1} \right)|q|^{1-\Gamma} + F_2|q|^{5-2\Gamma-z} = 0
\label{eq:Static5}\,,
\end{equation}
while Eq.~(\ref{eq:Herring4}) does not change and thus one still has $z=1$. However,  $\Gamma=1$ is no more a possible solution of Eq.~(\ref{eq:Static5}), since it is inconsistent with the defining conditions of this sector.  The remaining option is given by $1-\Gamma=5-2\Gamma-z$, which implies $\Gamma=3$. This solution corresponds to a roughness exponent of $\zeta=1$ and thus is inconsistent with the assumption of small gradients used in the derivation of the equations of motion. Therefore, possible solutions do not exist in sector $\beta$. In sector $\gamma$ defined by $(4-z)<\Gamma<(5-z)/2$, Fig.~\ref{fig:sectors} readily shows that one always has $\Gamma<1$ and thus physical solutions can not be found in this sector. Finally, in sector $\delta$ defined by $\Gamma>(4-z)$ and $\Gamma>(5-z)/2$, Eqs.~(\ref{eq:Static3},\ref{eq:Herring3}) are rewritten as
\begin{eqnarray}
 D + A_2 - A(1 - A_1)|q|^{1 - \Gamma} + (AF_1+F_2)|q|^{5-2\Gamma-z}&=& 0 \label{eq:Static7} \, , \\
 B|q|^z  - (1 - A_1)|q| + F_3|q|^{5-\Gamma-z} &=& 0 \label{eq:Herring6}.
\end{eqnarray}
From the defining conditions of this sector the last term in Eq.~(\ref{eq:Herring6}) is dominant over the second, which results in the scaling relation $z = (5-\Gamma)/2$. Plugging this relation into the defining condition of this sector, one finds that one would necessarily have $\Gamma>3$ (or equivalently $\zeta>1$), which is again inconsistent with the assumption of small gradients used to derive the equation of motion.

\subsection{The rough phase}
\begin{figure}[ht]
\centerline{\includegraphics[width=7cm]{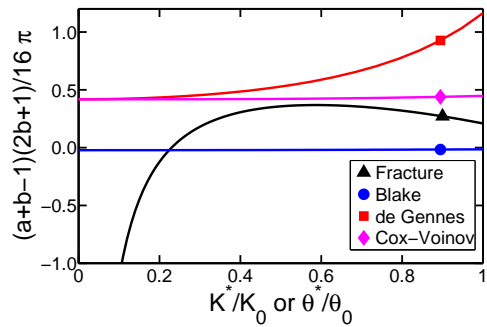}}
\caption{The existence of the rough phase is controlled by Eq.~(\ref{eq:roughcontrol}) and depends on the prefactor $(a+b-1)(2b+1)/(16 \pi)$ as a function of the dimensionless ratio $K^*/K_0$ (or $\theta_\mathrm{eq} / \theta_0$ for wetting) for the four different cases discussed in this paper. As can be seen, in three out of the four cases, the positivity of this prefactor is satisfied for low velocities, where its value is around $0.5$. For Blake's mobility law, this prefactor is always negative (yet very small) and for fracture it becomes negative at high velocities.}
\label{fig:a+b-1}
\end{figure}

Apparently, the only possibility one gets is the simple case $\Gamma = z = 1$ which corresponds to a moving flat front where perturbations propagate in a wave-like manner. However, a more careful inspection shows that one possibility was ignored, namely that of getting a fine-tuned case where the nonlocal elastic term is renormalized by the nonlinear term such that the pre-factor of $|q|^{1-\Gamma}$ vanishes in Eqs~(\ref{eq:Static4},\ref{eq:Static5},\ref{eq:Static7}). This means that one has to re-perform the analysis by taking into account this possibility, for which the higher order corrections given in Eqs.~(\ref{eq:integrals21}-\ref{eq:integrals23}) become important. In a previous publication~\cite{EPL}, the classical analysis in the different sectors has been repeated while taking into account this new instability. Here, we adopt a more direct approach, namely reconsidering the original SCE equations~(\ref{eq:Static2})-(\ref{eq:Herring2}). This approach clarifies the necessary conditions for the appearance of a rough phase. The idea is to expand $I_1(q), I_2(q)$ and $J(q)$ in powers of $q$ for small $q$'s
\begin{eqnarray}
 I_1(q)\simeq J(q) &\simeq& |q|\frac{(a+b-1)(2b+1)}{16\pi} \int\limits_0^\infty {\frac{\ell^3 \phi_\ell}{\omega_\ell} d\ell} \nonumber\\
 &+& |q|^2 \left({\frac{(a + b - 1)(2a - 1)}{32\pi} \int\limits_0^\infty  {\frac{\ell^3\omega'_\ell \phi_\ell}{\omega_\ell^2}d\ell} + \frac{{\left({1 + 2a + 2b - 2ab - 2a^2 } \right)}}{{16\pi }}\int\limits_0^\infty {\frac{\ell^2 \phi_\ell}{\omega_\ell} d\ell}} \right) + O(|q|^3) \label{eq:I1} \, , \\
 I_2(q) &\simeq& \frac{(a+b-1)^2}{16\pi} \int\limits_0^\infty {\frac{\ell^4 \phi_\ell^2}{\omega_\ell} d\ell} + O(|q|) \label{eq:I2} \, .
\end{eqnarray}
Interestingly $J(q)$ has the same expression as $I_1(q)$ up to
$2^{nd}$ order, which is the order that interests us here.
When plugging this into Eqs.~(\ref{eq:Static2})-(\ref{eq:Herring2})
we get
\begin{eqnarray}
D        &+& \frac{(a+b-1)^2}{16\pi} \int\limits_0^\infty {\frac{\ell^4 \phi_\ell^2}{\omega_\ell} d\ell} - |q|\left(1 - \frac{(a+b-1)(2b+1)}{16\pi} \int\limits_0^\infty {\frac{\ell^3 \phi_\ell}{\omega_\ell} d\ell}\right)\phi_q \nonumber \\
         &+&  \left({\frac{(a + b - 1)(2a - 1)}{32\pi} \int\limits_0^\infty  {\frac{\ell^3\omega'_\ell \phi_\ell}{\omega_\ell^2}d\ell} + \frac{{\left({1 + 2a + 2b - 2ab - 2a^2 } \right)}}{{16\pi }}\int\limits_0^\infty {\frac{\ell^2 \phi_\ell}{\omega_\ell} d\ell}} \right)|q|^2 \phi_q  = 0 \, , \label{eq:Static8}\\
\omega_q &-& \left(1 - \frac{(a+b-1)(2b+1)}{16\pi}\int\limits_0^\infty {\frac{\ell^3 \phi_\ell}{\omega_\ell}d\ell}\right)|q| \nonumber \\
         &+&\left({\frac{(a + b - 1)(2a - 1)}{32\pi}\int\limits_0^\infty  {\frac{\ell^3\omega'_\ell\phi_\ell}{\omega_\ell^2}d\ell} + \frac{{\left({1 + 2a + 2b - 2ab -2a^2 } \right)}}{{16\pi }}\int\limits_0^\infty {\frac{\ell^2\phi_\ell}{\omega_\ell} d\ell}} \right)|q|^2 = 0 \label{eq:Herring8} \, .
\end{eqnarray}
$I_2(0)$ only renormalizes the noise amplitude $D$ and it is
always a positive contribution. The analysis of the previous section is completely changed
whenever the prefactor of $|q|$ in the two equations above vanishes, i.e. whenever
\begin{equation}
\frac{(a+b-1)(2b+1)}{16\pi}
\int\limits_0^\infty {\frac{\ell^3 \phi_\ell}{\omega_\ell} d\ell} -1=0.\label{eq:roughcontrol}
\end{equation}
In this case the next order terms become important, and it immediately follows that $\phi_q \propto q^{-2}$ and $\omega_q\propto q^2$. Using previous notations, this means that $\Gamma=z=2$ and $\zeta=1/2$. Note however that in order to obtain this scenario
a necessary condition is that the quantity $(a+b-1)(2b+1)$ must be positive. In
Fig.~\ref{fig:a+b-1}, we plot this prefactor for the four cases discussed in this paper. Interestingly, in the case of wetting this prefactor is indeed positive for de Gennes' (Eq.~\ref{eq:JdG/PV}) and Cox-Voinov's (Eq.~\ref{eq:CV}) mobility laws. However for Blake's mobility law (Eq.~(\ref{eq:B})), this prefactor is negative so that no rough phase can exist. Also, in the case of fracture the propagation law (Eq.~\ref{eq:motionfrac}), imposes that  $(a+b-1)(2b+1)/(16 \pi)$ decreases with $K_0/K^*$ and even becomes negative for small $K_0/K^*$, and hence with increasing mean velocity $v$ of the front; the rough phase might disappear at high velocity of the elastic line.

Actually, since the possibility $\Gamma=z=2$ lies exactly on the boundary between the two sectors $\beta$ and $\delta$ (see Fig. \ref{fig:sectors}), integrals such as in Eq.~(\ref{eq:roughcontrol}) diverge logarithmically for small $q$. So the scenario must be slightly corrected as follows (see appendix B for details). The branch described by $\Gamma=z=1$ (the linear regime) rules for small $q$,  $q<q_\mathrm{c}$. Then from $q_\mathrm{c}$ up to $\Lambda$ (the upper cutoff introduced before), the scaling regime with $\Gamma=z=2$ is dominant. The transition point is given by
\begin{equation}
q_\mathrm{c} \sim \Lambda e^{-\frac{B}{A}\frac{16 \pi}{(a+b-1)(2b+1)}}\,,
\end{equation}
where $A$ and $B$ are defined by Eqs.~(\ref{eq:phiq})-(\ref{eq:omegaq}). The cutoff $q_\mathrm{c}$ is small as long as $(a+b-1)(2b+1)>0$, which is the same condition as that encountered above, so that the rough behaviour exists on a wide range $q_\mathrm{c}<q<\Lambda$. This justifies the statement that the opposite signs in the prefactor of $|q|$ in the equations can drive the spectrum towards a rough regime. Moreover, it is now clear that the requirement of Eq.~(\ref{eq:roughcontrol}) is not a strict one. It is sufficient to make this difference very small in order to get a small $q_\mathrm{c}$ (in other words $q_\mathrm{c} \rightarrow 0$ when the difference tends to zero) such that the rough branch of the spectrum takes over. We would expect then a structure factor of the form
\begin{equation}
\phi_q=\frac{A}{q_\mathrm{c}|q|+q^2}\,.\label{eq:strfactroughgen}
\end{equation}

\section{Discussion}

In this paper we derived from first principles the nonlinear equations of motion for an in-plane crack front, and for a wetting contact line using the three available local velocity laws.  These equations (\ref{eq:motionfrac},\ref{eq:B},\ref{eq:JdG/PV},\ref{eq:CV}) coupled with (\ref{eq:K},\ref{theta}) could serve as reference for future research.

Then we analyzed these equations with the aim of studying possible roughening of the front. To this end, we used the self-consistent expansion developed in~\cite{SCE} and found the possibility of having a rough moving phase with a roughness exponent $\zeta=1/2$ and a dynamic exponent $z=2$. These results are relevant when $K_0 \sim K^\star$ (or $\theta\sim\theta^\star$) and are due to destabilization of the nonlocal elasticity by the nonlinear terms. Including higher order terms would not affect these results as they have the same structure as the quadratic nonlinear terms which we took into account (see Eq.~\ref{theta}). Interestingly, since this destabilization is generic, it can be captured by other methods such as Dynamic Renormalization Group (DRG) used in~\cite{Golestanian}. For example, using our notations, the $1$-loop expansion of the full propagator (corresponding to Eq.~(61) in~\cite{Golestanian}) can be written as
\begin{equation}
 G^{-1}(q,0) = |q|\left( {1 - \frac{(a + b - 1)(2b + 1)}{16\pi}D\int_{}^\Lambda \ell d\ell} \right) + O\left(|q|^2\right)
\label{eq:DRG}\,,
\end{equation}
which resembles our result in Eq.~(\ref{eq:Static8}), having the same constant in front of the integral.
Again, $q^2$ relaxations in the system appear when the $|q|$-behaviour is killed by the nonlinear terms.

We found that this scenario was consistent with three out of the four equations of motions we considered, namely the equation for the crack front~(\ref{eq:motionfrac}) and the two equations for the wetting front using the mobility laws of de-Gennes~(\ref{eq:JdG/PV}) and Cox-Voinov~(\ref{eq:CV}). However, Blake's mobility law~(\ref{eq:B}) was found to exclude the rough phase. This finding could serve as an argument against Blake's mobility law \cite{blake} in the ongoing debate over the dynamical laws governing the spreading of fluids on solid substrates \cite{pomeaurev}.

Our finding is consistent with the roughness exponents measured in crack propagation, $\zeta \simeq 0.6$~\cite{Daguier,Schmittbuhl}, in water/glycerol contact lines, $\zeta\simeq 0.51$~\cite{Moulinet}. Our dynamic rough phase would apply to the quick motion of the lines in experiments. More precisely, a physical picture can be drawn as follows. In experiments, the driving force is not imposed at the front but mediated by a spring -- the meniscus for contact lines and the elastic medium for crack fronts. As a consequence, the force acting on the front varies in space and time; it oscillates below and above the local threshold for motion. Above this threshold, the front moves quickly (an avalanche occurs) and its roughness corresponds to our calculation. Then the force drops below the threshold and the front freezes. Here comes into play the discontinuity in the equation of motion (the Heaviside function) corresponding to the irreversibility of crack opening and to the hysteresis in equilibrium contact angle. This discontinuity becomes important in the last steps of freezing and tends to further roughen the line, with a roughness exponent slightly larger than $1/2$. Moreover, the roughness exponent  should be larger for the very irreversible crack propagation than for the slightly irreversible wetting,  as seen in experiments. To summarize, we propose that observations can be explained by a frozen dynamically rough interface.

Interestingly, a similar phenomenon is implicitly present in the KPZ system. It is well known that any rough surface would eventually flatten by the KPZ system if the noise is stopped \cite{EytanBD}. However in real situations, it is compensated by a non-zero ``angle of repose" that eventually freezes the system in a rough phase ~\cite{Moshe04} (this is expressed by an additional Heaviside function in the KPZ equation). It is also shown that the roughness exponent would be the same as that of the driven system if the freezing is done adiabatically \cite{Moshe04}. This shows that a hysteretic effect (existence of an angle of repose) can be consistently neglected once an above-threshold driving noise is present. It also hints that the final roughness of our system might be ``history dependent" i.e. it might depend on the protocol of the loading/freezing, if not done adiabatically.

Obviously, our study calls for more experimental and numerical work. In order to test our predictions, simulations such as in~\cite{Procaccia} could be performed for the nonlinear equations derived here and for a finite velocity of the elastic line. On the experimental side, several features could be tested: (i) the history dependance of the roughness exponent; (ii) the use of the form (\ref{eq:strfactroughgen}) to fit structure factors; (iii) the transition from a rough phase (roughness exponent around $1/2$) to a flat phase at high velocity. Another experimental challenge would be the direct measurement of the dynamic exponent $z$ predicted to be $2$ here, which would provide an additional test. An interesting possibility would be to investigate the relaxation of the contact line after applying spatially periodic perturbations at given wavelengths as done in~\cite{Ondar}.

\section*{Acknowledgements}
This work was supported by EEC PatForm Marie Curie action (E.K.). We thank D. Vandembroucq for fruitful discussions. Laboratoire de Physique Statistique is associated with Universities Paris VI and Paris VII.

\section*{Appendix A}

In this appendix, we recall some elements of the elastodynamic theory of cracks~\cite{Freund} and apply them to the motion of slow cracks. We begin with the definition of the Rayleigh wave speed $c_\mathrm{R}$, which is the root of $D(v)$ given by
\begin{eqnarray}
D\left( v \right) = 4\alpha _d \alpha _s  - \left( {1 + \alpha _s^2 } \right)^2  = 0
 \label{eq:D(v)} \, ,\\
\textrm{with}\quad \alpha _{s,d}  = \sqrt {1 - \tfrac{{v^2 }} {{c_{s,d}^2 }}}
 \label{eq:alphasd} \, ,
\end{eqnarray}
where $c_\mathrm{d}$ and $c_\mathrm{s}$ are the dilatational and shear wave speeds. In the expression for the dynamic energy release rate $G$ in
Eq.~(\ref{eq:Griffith}), we used the universal function~\cite{Freund}
\begin{equation}
g \left( v \right) = \frac{\left(1 - \alpha _s^2\right)\alpha _d}{D\left( v \right)}\frac{\left(1 - v /c_\mathrm{R} \right)^2 }{1- v/c_d}
\exp\left\{ - \frac{2}{\pi }\int\limits_1^{\sqrt \kappa  } {\arctan
\left[ {\frac{{4\eta ^2 \sqrt {\left( {\eta ^2  - 1} \right)\left(
{\kappa  - \eta ^2 } \right)} }}{{\left( {\kappa  - 2\eta ^2 }
\right)^2 }}} \right]\frac{{d\eta }}{{\eta  - c_d/v}}} \right\}
 \label{eq:defg1} \, ,
\end{equation}
where $\kappa=(c_d/c_s)^2$. To derive the nonlinear equation of motion, we need an expansion of the function $\sqrt{g(v)}$ to second order in $v$, which can be written in full generality as
\begin{equation}
\sqrt {g (v)}  \simeq 1 - \alpha \left( \kappa
\right)\frac{v}{{c_\mathrm{R} }} +\beta(\kappa)\left( {\frac{v}{{c_\mathrm{R} }}} \right)^2  +
\mathcal{O}\left( \left(\frac{ v}{c_\mathrm{R}} \right)^3 \right) \, .
\end{equation}
After algebraic manipulations, it can be shown that the coefficients $\alpha(\kappa)$ and $\beta(\kappa)$ can be explicitly written as
\begin{eqnarray}
\alpha \left( \kappa  \right)& = &1 - \frac{1}{2}\frac{c_\mathrm{R}}{c_d} - \frac{{c_\mathrm{R}}}{c_d}\int\limits_1^{\sqrt \kappa  } {\arctan \left[ {{\textstyle{{4\eta
^2 \sqrt {\left( {\eta ^2  - 1} \right)\left( {\kappa  - \eta ^2 }
\right)} } \over {\left( {\kappa  - 2\eta ^2 } \right)^2 }}}}
\right]\frac{d\eta}{\pi} }
 \label{eq:alpha} \, ,\\
\beta \left( \kappa  \right) &\equiv&  \frac{1}{2}\alpha^2(\kappa)
 \label{eq:beta} \, ,
\end{eqnarray}
which in turn yields Eq.~(\ref{eq:g1}).

\section*{Appendix B}

In this appendix we find an estimation of $q_\mathrm{c}$, the transition scale between the rough branch described by $\Gamma=z=2$ and the (logarithmically rough) linear branch described by $\Gamma=z=1$.
Our starting point are Eqs.(\ref{eq:Static8}-\ref{eq:Herring8}), from which we get
\begin{eqnarray}
 \phi_q   &=& \frac{\hat D} {\left(1 - W_1\right)|q| + W_2|q|^2} \label{eq:phiq2} \, , \\
 \omega_q &=& \left(1 - W_1\right) |q| + W_2 |q|^2  \label{eq:omegaq2} \, ,
\end{eqnarray}
with
\begin{eqnarray}
 \hat D &\equiv& D + \frac{(a + b - 1)^2}{16\pi} \int\limits_0^\infty {\frac{\ell^4 \phi_\ell^2}{\omega_\ell} d\ell} \, , \\
 W_1    &\equiv& \frac{(a + b - 1)(2b + 1)}{16\pi} \int\limits_0^\infty {\frac{\ell^3 \phi_\ell}{\omega_\ell} d\ell} \, , \\
 W_2    &\equiv& \frac{(a + b - 1)(1 - 2a)}{32\pi} \int\limits_0^\infty {\frac{\ell^3 \omega'_\ell \phi_\ell}{\omega_\ell^2} d\ell} - \frac{(1 + 2a + 2b - 2ab - 2a^2)}{16\pi} \int\limits_0^\infty {\frac{\ell^2 \phi_\ell}{\omega_\ell} d\ell} \, .
\end{eqnarray}
From the equations above, we expect that at low $q$, $\omega_q = \left| q \right|$ and $\phi_q=D |q|^{-1}$, which corresponds to the linear branch; whereas at larger $q$, $\omega_q = B|q|^2$ and $\phi_q = A|q|^{-2}$, which corresponds to the rough branch (see Eqs.~\ref{eq:phiq}-\ref{eq:omegaq}). The transition between the two behaviors occurs when $q=q_\mathrm{c}$ determined by $1-W_1=W_2q_\mathrm{c}$. Using the small scale cutoff defined by $\Lambda$, which was introduced in Eqs.~(\ref{eq:A1},\ref{eq:A2}), we can estimate
\begin{eqnarray}
 \int\limits_0^\infty {\frac{\ell^3 \phi_\ell}{\omega_\ell}d\ell} &=& \int\limits_0^{q_\mathrm{c}} {\frac{\ell^3 \phi_\ell}{\omega_\ell} d\ell}  + \int\limits_{q_\mathrm{c} }^\Lambda {\frac{\ell^3 \phi_\ell}{\omega_\ell}d\ell} + \int\limits_\Lambda^\infty {\frac{\ell^3 \phi_\ell}{\omega_\ell}d\ell}  \\
 &=& D\int\limits_0^{q_\mathrm{c}} {\ell d\ell } + \frac{A}{B}\int\limits_{q_\mathrm{c}}^\Lambda {\frac{d\ell}{\ell}} + \int\limits_\Lambda ^\infty  {\frac{{\ell ^3 \phi _\ell  }}{{\omega _\ell  }}d\ell }  = {\textstyle{1 \over 2}}Dq_\mathrm{c}^2  + \frac{A}{B}\log \frac{\Lambda }{{q_\mathrm{c} }} + \int\limits_\Lambda ^\infty  {\frac{{\ell ^3 \phi _\ell  }}{{\omega _\ell  }}d\ell} \, ,
\end{eqnarray}
and deal with similar integrals in the same way. Hence the defining equation of $q_\mathrm{c}$ becomes
\begin{eqnarray}
&& \frac{{\left( {a + b - 1} \right)\left( {2a -
1} \right)}}{{32\pi }}\left( {Dq_\mathrm{c}  + 2\frac{A}{B}\left(
{\frac{1}{{q_\mathrm{c} }} - \frac{1}{\Lambda }} \right) +
\int\limits_\Lambda ^\infty  {\frac{{\ell ^3 \omega '_\ell  \phi
_\ell  }}{{\omega _\ell ^2 }}d\ell } } \right)\nonumber\\
&+& \frac{{\left( {1 +
2a + 2b - 2ab - 2a^2 } \right)}}{{16\pi }}\left( {Dq_\mathrm{c}  +
\frac{A}{B}\left( {\frac{1}{{q_\mathrm{c} }} - \frac{1}{\Lambda }} \right) +
\int\limits_\Lambda ^\infty  {\frac{{\ell ^2 \phi _\ell  }}{{\omega
_\ell  }}d\ell } } \right) q_\mathrm{c} \nonumber \\
&=& {1 - \frac{{\left( {a
+ b - 1} \right)\left( {2b + 1} \right)}}{{16\pi }}\left(
{{\textstyle{1 \over 2}}Dq_\mathrm{c}^2  + \frac{A}{B}\log \frac{\Lambda
}{{q_\mathrm{c} }} + \int\limits_\Lambda ^\infty  {\frac{{\ell ^3 \phi _\ell
}}{{\omega _\ell  }}d\ell } } \right)}  \, .
\end{eqnarray}
In the limit $W_1\to 0$, we get $q_\mathrm{c}\to 0$, so that we can keep the most dominant terms and get
\begin{equation}
q_\mathrm{c}  = \Lambda \exp\left(-\frac{B}{A}\frac{16\pi}{(a + b - 1)(2b + 1)} + \frac{(a + b + 2)}{(a + b - 1)(2b + 1)}\right) \, .
\label{eq:qc}
\end{equation}

\end{document}